\documentclass[10pt,conference]{IEEEtran}
\usepackage{amsmath, amssymb, bm, cite, epsfig, psfrag}
\usepackage{graphicx}
\usepackage[margin=0.7in]{geometry}
\usepackage{dblfloatfix}
\usepackage{array}
\newcolumntype{P}[1]{>{\centering\hspace{0pt}}p{#1}}
\newcolumntype{M}[1]{>{\centering\hspace{0pt}}m{#1}}
\newcolumntype{L}{>{\centering\arraybackslash}m{3cm}}
\usepackage{longtable}
\usepackage{supertabular,booktabs}
\usepackage{multirow}
\usepackage[usenames,dvipsnames]{xcolor}

\usepackage{etoolbox}
\usepackage{pbox}
\usepackage{fixltx2e}
\usepackage{filecontents}
\usepackage{enumerate}
\graphicspath{ {figures/} }
\usepackage{subfigure}
\usepackage{colortbl}
\usepackage{fancyhdr}

\usepackage{bm}
\newtoggle{conference}
\togglefalse{conference} %
\graphicspath{{figures/}}

\fancypagestyle{firststyle}{
	\fancyhf{}
	\fancyhead[L]{O. Kanhere and T. S. Rappaport, ``Position Locationing for Millimeter Wave Systems,'' \textit{in GLOBECOM 2018 - 2018 IEEE Global Communications Conference,} Abu Dhabi, U.A.E., Dec. 2018, pp. 1--6.}     %

}
\usepackage{etoolbox}
\makeatletter
\patchcmd{\@makecaption}
{\scshape}
{}
{}
{}
\makeatletter
\patchcmd{\@makecaption}
{\\}
{.\ }
{}
{}
\makeatother

\begin{document}
\title{Position Locationing for Millimeter Wave Systems }
\author{\IEEEauthorblockN{Ojas Kanhere, Theodore S. Rappaport\\}
	
\IEEEauthorblockA{	\small NYU WIRELESS\\
					NYU Tandon School of Engineering\\
					Brooklyn, NY 11201\\
					\{ojask,tsr\}@nyu.edu}}\vspace{-0.7cm}

\maketitle
\thispagestyle{firststyle}

\begin{abstract}
The vast amount of spectrum available for millimeter wave (mmWave) wireless communication systems will support accurate real-time positioning concurrent with communication signaling. This paper demonstrates that accurate estimates of the position of an unknown node can be determined using estimates of time of arrival (ToA), angle of arrival (AoA), as well as data fusion or machine learning. Real-world data at 28 GHz and 73 GHz is used to show that AoA-based localization techniques will need to be augmented with other positioning techniques. The fusion of AoA-based positioning with received power measurements for RXs in an office which has dimensions of 35 m by 65.5 m is shown to provide location accuracies ranging from 16 cm to 3.25 m, indicating promise for accurate positioning capabilities in future networks. Received signal strength intensity (RSSI) based positioning techniques that exploit the ordering of the received power can be used to determine rough estimates of user position. Prediction of received signal characteristics is done using 2-D ray tracing.

\end{abstract}
    
\begin{IEEEkeywords}
    positioning; position location; navigation; mmWave; 5G; ray tracing; site-specific propagation;
\end{IEEEkeywords}

\section{Introduction}
The global positioning system (GPS) has become a ubiquitous technology, allowing users to find their location in unknown places nearly anywhere in the world. However, not all locations are adequately covered. Urban canyons with high rise buildings and indoor locations suffer from poor GPS accuracy, although averaging and enhancements such as assisted GPS (AGPS)\cite{Djuknic01a} and differential GPS (DGPS)\cite{Kaplan05a} have provided improved locationing. 

AGPS provides the user device with orbital data of GPS satellites, narrowing the search window where the receiver (RX) must search for the satellite's signal and allowing the GPS RX to dwell for longer times in different time and frequency bins. When the dwell time in each bin is increased by a factor of ten, the RX sensitivity increases by 10 dB\cite{Diggelen02a}. In addition, since ultra high frequency (UHF)/microwave cellular signals can penetrate through building walls, AGPS helps in indoor positioning. As mmWave cellular networks proliferate over the next decade, building penetration will become more difficult \cite{Rap13a}, and new methods will be needed to assist with positioning.

GPS RXs estimate the distance to satellites using time of flight of the signals. Distance estimates are called the pseudo-ranges of the satellites. GPS RXs separated by distances of less than 100 km have similar pseudo-range errors\cite{Kee90a}. A DGPS RX station (RS) is located at a position known to the RS. The RS calculates the difference between the pseudo-ranges and actual distances to all visible satellites. The RS then transmits these differences as corrections to all end users in the RS's coverage area. The end user selects the corrections corresponding to the visible satellites and then subtracts these corrections from the end user's measured pseudo-ranges. Using DGPS, the positioning accuracy achieved by users is 50 cm in the vicinity of the RS. The accuracy degrades by 20 cm every 100 km away from the DGPS RS \cite{monteiro05}.

Accurate positioning systems are required for a wide variety of applications such as guided museum tours \cite{Kolodziej06a}, navigation in large malls \cite{Puikkonen09a} or office spaces, see-in-the-dark capabilities for firefighters and law enforcement, IoT device tracking \cite{Zhang12a}, personnel tracking, and accurate positioning of equipment for automation in smart factories \cite{Rap89c,Lu17}.

This paper is organized as follows: Section \ref{section2} provides an overview of positioning techniques and explains how the large bandwidth of future 5G and 6G mmWave communication systems operating at frequencies above 100 GHz \cite{Ychou18a} enables accurate positioning. Section \ref{section3} describes time and phase shift based positioning methods. In sections \ref{section6} and \ref{section7}, the AoA-based and RSSI vector-based positioning methods are described and the performance of both techniques is evaluated using real-world data at mmWave frequencies of 28 and 73 GHz \cite{Deng15a,Mac17f}. The fusion of AoA-based positioning with received power measurements for RXs in an office which has dimensions of 35 m $ \times $ 65.5 m is shown to provide location accuracies ranging from 16 cm to 3.25 m, indicating promise for accurate positioning capabilities in future networks. In section \ref{section8} a 2-D ray tracer is introduced which is calibrated using indoor mmWave measurement data at 28 GHz. The paper concludes with suggestions for further work.

\section{Positioning in Future mmWave Wideband Communication Systems}\label{section2}
The problem of positioning in communication systems involves finding the location of a user, given the known locations of other stations, that may be mobile or fixed. There are several well-known methods that can be used to determine the position of the user. Positioning methods using ToA\cite{Rap96a} rely on computations of the radio propagation time delays (or distances, which are directly related to the propagation speed of radio waves in free space) between different stations with known or unknown positions from the transmitter (TX) to reach the RX. ToA-based positioning requires tight time synchronization between the TX and RX. To measure the time difference of arrival (TDoA) for positioning, synchronization between TX nodes is required. Based on TDoA measurements between a pair of TXs, the RX can be localized to one arm of a hyperbola \cite{Rap96a}. By using two pairs of TXs, the position of the RX can be determined to be the point that is at the intersection of the two hyperbolas, as seen in Fig. \ref{fig:TDoA}.
 
AoA-based positioning techniques estimate the line of bearing based on the spatial directional received power at the RX, and is well suited for narrow beam steerable antennas that can scan to find the AoAs of three or more sources with known locations (the ``three point problem") \cite{Rappaport88a}. Using geometric principles, the line of bearing to multiple TXs are used to localize the RX \cite{Rappaport88a,Rap96a} using knowledge of the RX AoAs from various beacons. AoA-based positioning was proven effective in factory environments for autonomous vehicles that used a scanning optical RX\cite{Rappaport88a}.

Using RSSI-based positioning approaches, the RX can be positioned, since RSSI generally monotonically decreases with distance \cite{Bahl00}. The RSSI can be determined for a known environment (e.g. a building) using pre-measured signal strengths and known TX locations over a grid of locations (e.g. a map of the environment) from multiple TXs. By applying a suitable path loss model or pre-measured values over a grid of locations, with fine enough resolution to offer the desired accuracy in positioning, the distance between the TX and RX can be estimated \cite{Bahl00}. For many different TXs with known locations, the resulting position is estimated based on the best fit, e.g. the most likely location that best matches all measured signal levels from the multiple TXs \cite{Youssef05}.

The mmWave regime offers much wider bandwidths than has ever been available to mobile networks \cite{Rap14a}, thus offering both lower latency, and a built-in position location capability due to the massive transmission bandwidth. A passband signal with bandwidth $ B $ Hz that is sampled at its Nyquist rate must be sampled every $ 1/B $ seconds at baseband. The raw resolution of mmWave systems, defined as the distance light travels between these sampling instants, is the smallest distance difference that can be measured. With available bandwidth $ B $, the raw resolution is given by \cite{Lem16a}:

\begin{align}
	r = \dfrac{c}{B},\label{raw_res}
\end{align}
where $r$ is the raw resolution in meters, $c$ is the speed of light in meters per second.

The raw resolution of mmWave systems is much smaller than the resolution of systems operating at sub-6 GHz frequencies owing to the much greater bandwidths available.  LTE systems use the position reference signal (PRS) for positioning, which is a pseudo-random sequence sent by LTE base stations \cite{3GPP_36_355}. The 20 MHz 4G LTE channel can use a PRS that has a bandwidth of at most 20 MHz \cite{3GPP_36_355} and thus has a raw resolution of 15 m. A 4 GHz bandwidth mmWave or Terahertz (THz) channel could provide a raw resolution of 7.5 cm. Better accuracy could be obtained by fusing ToA with accurate AoA information, available due to the narrow beamwidths of mmWave antennas, and RSSI.

\section{Time and Phase Based Localization methods} \label{section3}
An anchor node is a node whose position is known by the user. Since electromagnetic waves propagate at the speed of light, the one-way (e.g. from anchor to user) propagation time can be used to estimate the distance of the user from an anchor node using the relation:
\begin{align}
d = c \times t,\label{distance_from_time}
\end{align}	
where $d$ is the estimated distance of the user from the anchor node and $t$ is the one-way propagation time. 

To make the distance estimate accurate, tight synchronization between the user and the anchor node is necessary. Nanosecond-level timing synchronization is required for sub-meter accuracy, which is costly and increases the system complexity. 

Simpler systems that do not need synchronization can be designed which measure roundtrip propagation time instead \cite{Mao07}. The user and the anchor node estimate their signal processing time, which is then be calibrated out to give the round-trip time equal to twice the one-way propagation time.

The TDoA method determines the location of the user relative to the anchor nodes by measuring the difference in arrival times of signals from the anchor nodes. By measuring the difference in the ToA of signals from two pairs of anchor nodes, the difference in distance can be calculated using (\ref{distance_from_time}).
\begin{figure}[]
	\centering
	\includegraphics[width=0.4\textwidth]{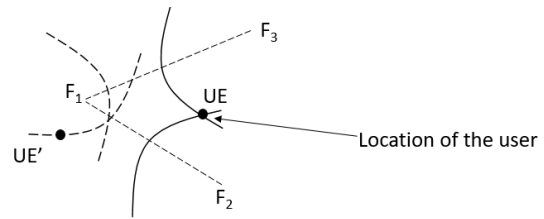}
	\caption{The UE is located at a point that is at the intersection of the arms of hyperbolas with foci at anchor nodes ($F_1$, $F_2$) and ($F_1$, $F_3$), that correspond to a difference in distance of $ k $.  UE$'$ lies on the arm of the hyperbola corresponding to a difference in distance of $-k $ \cite{Rap96a}. }
	\label{fig:TDoA}
\end{figure}
\begin{figure}[]
	\centering
	\includegraphics[width=0.4\textwidth]{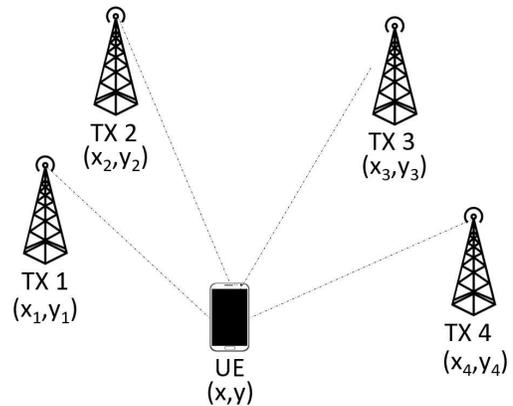}
	\caption{The UE is in range of several anchor nodes (TXs), which have known coordinates. The UE's coordinates are determined from different characteristics of the signals that arrive at the user from the anchor nodes. }
	\label{fig:pg1_base}
\end{figure}
If the difference in distance of the  user equipment (UE) from two anchor nodes, $F_1$ and $F_2$, is fixed, i.e. if $ d_{UE-F{_1}} - d_{UE-F{_2}} = k, $ the locus of the UE is one arm of a hyperbola with foci at  $F_1$ and $F_2$ (the other arm of the hyperbola corresponds to when the difference is $-k$, as seen in Fig. \ref{fig:TDoA}). The location of the user is the point that is at the intersection of the two hyperbolas, as seen in Fig. \ref{fig:TDoA}. 

The cross-correlation of two received signals can be used to measure the TDoA. The delay that maximizes the cross-correlation function is the TDoA \cite{Rap96a}.

A linear frequency modulated (LFM) signal, also called a chirp signal, has an instantaneous frequency which varies linearly with time over the chirp duration $ T_c $. These signals can be used to estimate the TDoA \cite{Huang08a,Adib15a}. At the RX, received signals from two TXs are mixed and then passed through a low pass filter to give a low frequency signal $ r(t) $. The instantaneous frequency of $ r(t) $ is proportionate to the TDoA of the two received signals \cite{Huang08a}. Thus, the problem of estimating the TDoA reduces to estimating the instantaneous frequency of $ r(t) $.

An  estimate of the phase accrued in the received signal can also be used to position the RX. The phase of the complex wireless channel, $ \angle h $, is $ 2 \pi $ periodic. For the RX locations with ToAs that differ by integral multiples of $1/f$, $ \angle h $ is the same. The ToA of an RX can be resolved up to an integral multiple of $ 1/f $. This ambiguity can be resolved by taking measurements at multiple frequencies, $ f_1, f_2,  ... , f_n$, in which case the ToA of an RX can be resolved up to an integral multiple of the least common multiple (LCM) of $ (1/f_1, 1/f_2 , ... ,  1/f_n ) $ \cite{Vas16a}.

\section{AoA-Based Localization} \label{section6}
A mobile user can be located using the AoAs of signals from a set of anchor nodes. The position can be determined to be the point that is at the intersection of the line of bearing from the anchor nodes to the user. Two such lines of bearing are sufficient to determine the location of the user.

Ideally, in noise-free environments, all lines of bearing would intersect at a single point. However, due to measurement errors and system noise, this is practically never the case when there are more than two lines of bearing \cite{Rappaport88a}. The user's location can then be estimated to be the least square solution for the intersection of the lines.

To determine the AoAs of the signal from each anchor node, the user's antenna should be mechanically or electrically rotated, until the signal with maximum power is received \cite{Wang17a}. In mmWave systems, this direction can be determined during initial access\cite{Giordani16a} or in various antenna search or handoff methods likely to evolve as part of standards that incorporate directional antennas \cite{Wang17a}.

\subsection{Accuracy of AoA-Based Positioning for indoor mmWave systems}\label{AoA_A}
\begin{figure*}
	\centering     %
	\subfigure[Map of the 9th floor of 2 MTC showing the TX and RX locations where the indoor measurements were conducted at 28 GHz and 73 GHz\cite{Deng15a}. The blue lines show the propagation paths predicted by 2D ray tracing, from TX 1 to RX 6.]{\label{fig:Indoor_locations}\includegraphics[width=93mm]{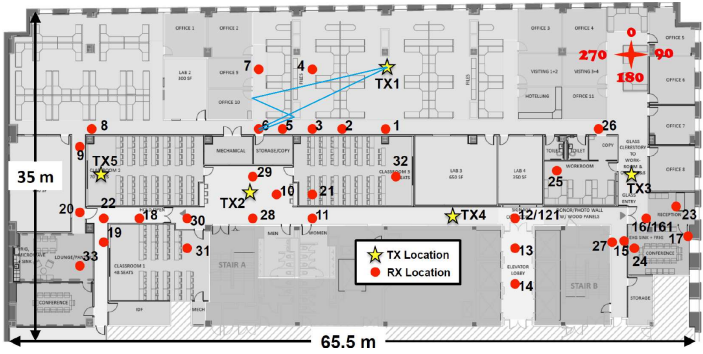}}
	\hfill
	\subfigure[The TX and RX locations for the outdoor base station diversity propagation measurements at 73 GHz\cite{Mac17f}.]{\label{fig:ap_diversity_locations}\includegraphics[width=60mm]{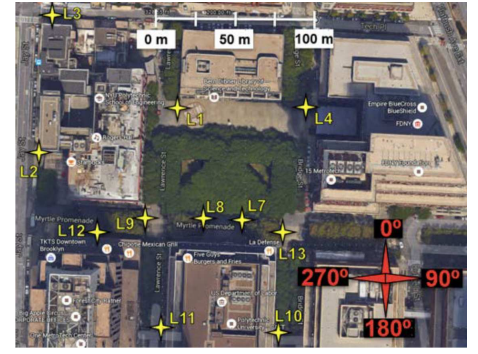}}
	\caption{Maps of the indoor and outdoor environments  where the AoA-based positioning technique \cite{Rap96a} was tested.}
\end{figure*}

To investigate the accuracy of various position location methods, measured data from a 2014 measurement campaign at 28 and 73 GHz, at the NYU WIRELESS research center\cite{Deng15a} was used. See Fig. \ref{fig:Indoor_locations} for the indoor environment. A 400 Megachip-per-second (Mcps) wideband sliding correlator channel  sounder with high-gain steerable antennas was used to measure the propagation channel. Five TX and 33 RX locations were chosen for a total of 48 TX-RX pairs. Out of the 33 RX locations, 11 were in the range of at least two TXs (only RX 16 was in the range of three TXs).

Six AoA sweeps and two angle of departure (AoD) sweeps were conducted at each RX location. The lines of bearing of the TX locations were determined to be the azimuth directions with maximal received power. The angular sweeps were centered around the angle at which maximum power was received, allowing lines of bearing to be measured to within one degree (the step size of the gimbal). Sweeps were done in increments of the half power beam width (HPBW) of the antennas which was $15^\circ$ at 28 GHz and $ 30^\circ $ at 73 GHz. The power delay profile (PDP) was recorded by time-averaging 20 consecutive PDPs \cite{Deng15a}.

It was observed that for RXs in NLOS environments, the strongest arriving beam was not the direct beam. Strong reflections were observed off walls and metallic surfaces. Hence the AoA-based method could not be applied to accurately localize such nodes. RX 8, 11, 13, 18 and 28 were in LOS of one TX. The TXs for each RX are listed in Table \ref{tab:table1}. The mean positioning error for the five LOS nodes was found to be poor: 11.01 m at 28 GHz and 6.20 m at 73 GHz, due to the absence of two LOS TXs. Accurate positioning could be done if the locations of the reflections, themselves, were ascertained by the RX, when a map of the environment is available \cite{WangH05}. This could be done through ray tracing as described in section \ref{section8}. Alternatively, angular information could be used for machine learning based fingerprinting, as described in section \ref{4C}.
\begin{table}[]
	\begin{center}
		\caption{TX locations used to transmit to each indoor RX location \cite{Deng15a}.}
		\label{tab:table1}
		\begin{tabular}{c|c} %
			\textbf{RX Location} & \textbf{Serving TX Locations} \\
			\hline
			 1 & 1\\
			 4 & 1\\
			 7 & 1\\		
			 8 & 1, 5 \\
			 10 & 2\\
			11 & 2, 4\\
			12 & 2, 4\\
			13 & 2, 4\\
			16 & 2, 3, 4\\
			18 & 2, 4\\
			28 & 4, 5
		\end{tabular}
	\end{center}

	\begin{center}
		\caption{TX locations used to transmit to each outdoor RX location \cite{Mac17f}.}
		\label{tab:table2}
		\begin{tabular}{c|c} %
			\textbf{RX Location} & \textbf{Serving TX Locations} \\
			\hline
		L1          & L3, L4, L7, L11, L13 \\
		L2          & L3, L9, L12          \\
		L4          & L1, L3, L7, L10, L13 \\ 
		L7          & L1, L2, L4, L10      \\
		L8          & L1, L7, L9           \\
		L9          & L1, L2, L4, L11      \\
		L10         & L4, L7, L13          \\
		L12         & L1, L2, L4, L7, L11  \\
		L13         & L1, L4, L10          
		\end{tabular}
	\end{center}
	\vspace{-20pt}
\end{table}

The AoA-based positioning algorithm \cite{Rap96a} was also tested on outdoor propagation data. In the summer of 2016, an outdoor measurement campaign was conducted on the NYU Tandon campus in Brooklyn, New York in an open-square (O.S.) environment, to study base station diversity \cite{Mac17f}. Measurements of the outdoor propagation channel were conducted at 73 GHz using a 500 Mcps wideband sliding correlator channel sounder with high-gain steerable antennas. Nine RX locations were in the coverage area of three to five TXs. The TXs for each RX are listed in Table \ref{tab:table2}. Of these nine RX locations, the positioning error at RX 1, 8, 9, 12, and 13 was less than 40 m, but not particularly good due to absence of true LOS environments at each RX.

\subsection{Combining AoA and Path Loss Information for Greater Accuracy}
The low accuracy of the AoA-based positioning method suggests that in mmWave systems, the AoA of strongest received signal does not necessarily correspond to the AoA of the LOS signal. In order to achieve better accuracy, we considered a fusion of multiple positioning methods.

Consider RX locations in the LOS of at least one TX location. Based on the power received by the RX, its distance from the LOS TX can be estimated using a suitable path loss model \cite{Deng15a} for the environment. Based on the  1 m close-in free space CI path loss model \cite{Sun16c}, the maximum likelihood (ML) distance $ d_{ML} $ is:
\begin{align}
d_{ML} [m] = d_0 [m] \cdot 10^{\left( 	PL [dB] -PL_{FS}(d_0) [dB] \right)/10 \cdot \bar{n} }
\end{align}
where $ d_0 $ = 1 m, $ PL $ is the path loss measured at the RX, $ PL_{FS}(d_0) $ is the free space path loss at $ d_0 $, and $  \bar{n} $ is the directional path loss exponent. In LOS environments, at 28 GHz and 73 GHz, $ \bar{n}=1.7 $\cite{Deng15a}.

\begin{figure}[]
	\centering
	\includegraphics[width=0.4\textwidth]{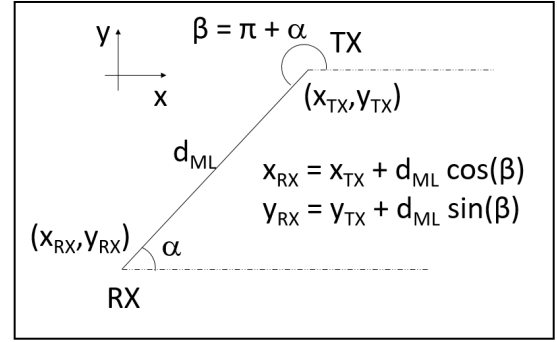}
	\caption{The coordinates of the RX can be determined based on the position of a TX in LOS, the TR separation distance and bearing angle between the TX and RX.}
	\label{fig:fusion}
\end{figure}

The AoA $ \alpha $ is assumed to be the antenna pointing angle at which the RX receives the maximum power (this assumption fails in the presence of strong specular reflection or in NLOS). Since the TX and RX are in LOS, the AoD, $ \beta= \alpha + \pi$. Given the bearing angle and TR separation distance, the position of the RX can be determined, as shown in Fig. \ref{fig:fusion}.

\begin{figure}[]
	\centering
	\setlength{\belowcaptionskip}{-0.5 cm}
	\includegraphics[width=0.4\textwidth]{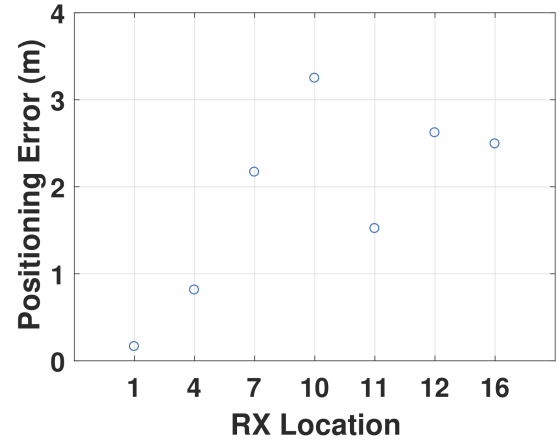}
	\caption{ Positioning errors for 28 GHz and 73 GHz indoor locations using a combination of AoA and received power measurements.}
	\label{fig:fusion_errors}
	\vspace{-15pt}
\end{figure}

The fusion of AoA and path loss based positioning methods gave results with much greater accuracy for indoor LOS locations at 28 GHz and 73 GHz compared to positioning using the AoA alone. See Fig. \ref{fig:Indoor_locations} for the indoor environment. Eight RXs that were served by one LOS TX were localized using this method. After removing one outlier (RX 28), the mean positioning error of the remaining seven nodes was 1.86 m with a minimum positioning error of 16 cm and a maximum positioning error of 3.25 m as seen in Fig. \ref{fig:fusion_errors}. The TXs for each RX are listed in Table \ref{tab:table1}.

\subsection{Machine Learning Based Fingerprinting}\label{4C}
We propose a machine learning based positioning technique for mmWave systems which could further increase positioning accuracy. Conventional fingerprinting techniques map the RSSI throughout the environment in consideration\cite{Bahl00}. We propose to also measure the AoA and ToA to increase redundancy and improve resilience to changes in the environment since such capabilities will naturally be available in wideband, narrowbeam mmWave and THz systems.

RSSI, AoA, and ToA are measured over a grid of locations from multiple TXs. These measured parameters are then stored in a database, off-line. Using these pre-measured signals, the user's position is estimated to be the most likely location that best matches all measured parameter values. Owing to the fine resolution of ToA measurements possible due to the wide bandwidth available, centimeter level positioning accuracy could be achieved, and we are presently exploring this approach. 

\section{Localization based on Connectivity and RSSI Rank }\label{section7}
A grid-based localization algorithm for sensor networks was proposed in \cite{Wang18a}. Each user ranks all the anchor nodes based on RSSI. Ideally, RSSI values decrease monotonically with distance. Hence, by ordering the RSSI values, the user can order the anchor nodes based on their distance. In a sensor network with $ m $ anchor nodes, the ordering is stored in an $m$-dimensional vector called the distance rank vector.

\subsection{The Distance Rank Vector}
The 2D plane can be subdivided into sub-regions based on the ordering of distances of the user from anchor nodes. All points within a sub-region have the same ordering and hence have identical distance rank vectors. Thus, the user can be localized to one of the many sub-regions in the plane based on the user's RSSI rank vector \cite{Wang18a}.

When there is an RSSI measurement error (or if RSSI does not monotonically decrease with distance due to multipath or obstructions/reflections), the RSSI vector may produce an infeasible distance rank vector. In this case, the user is localized to the sub-region whose distance rank vector is closest to the measured distance rank vector. The closeness of two distance rank vectors can be measured using the Spearman rank order correlation coefficient, $\rho $ $\in$ [-1,1], with close distance rank vectors having greater values of $\rho$ \cite{Wang18a}. 
\subsection{Grid-based Localization}
For $m$ anchor nodes, each user has $O(m^5)$ worst-case space and $O(m^5\log_2m)$ worst-case time complexity to create the distance rank vector \cite{Wang18a}. For resource-constrained user nodes, real time computation may not be possible. In such cases, a grid-based approach can be used.

A region called the \textit{estimation rectangle} (ER) is defined, surrounding each anchor node, based on an estimate of the anchor node's communication range. The ER is then divided into square-shaped grids, based on a predefined grid size. Within the area of intersection of ERs of all anchor nodes belonging to the user's distance rank vector, the set of grids with maximal Spearman correlation coefficient is called the \textit{residence area} of the user. The user's position is estimated to be the centroid of all grids within the residence area \cite{Wang18a}.

\subsection{Performance of the RSSI Rank-Based Algorithm Using Outdoor Measurement Data}
The RSSI rank-based algorithm was tested on the base-station diversity data described in \cite{Mac17f} which was obtained using the system described in section \ref{AoA_A}. The maximum range of communication was taken to be 200 m, the typical cell radius of a small-cell mmWave deployment. The TXs for each RX are listed in Table \ref{tab:table2}. The average positioning error was 34 m for grids with a grid length of 20 m. There was no accuracy improvement when the grid length was further reduced. Of the nine RX locations, L1, L2, L7, L9 and L10 had a positioning error less than 20 m.%

The mean distance error for the RSSI rank-based algorithm is quite large when there are limited known nodes with large coverage areas. However, the RSSI rank-based algorithm successfully localized the user to the correct sub-region, which could be used to determine the room or building location the user is in by positioning the anchor nodes so that sub-regions coincide with room or building boundaries.

\section{Ray Tracing of mmWave Communication Channels}\label{section8}
Real-world mmWave measurements are time intensive and expensive. In order to accurately predict received signal characteristics such as received power, needed to investigate position location algorithms, AoA and ToA, ray tracing may be used, so long as the ray tracer is extremely accurate and calibrated to observable measurements \cite{Rap15a,Seidel94a}. To explore likely position location accuracy, a 2-D ray tracer has been developed for this paper. 

Ray tracing was done by brute force, by launching 100 rays uniformly over $360^\circ$ in the azimuth plane at zero degree elevation from TX locations previously measured in \cite{Deng15a}. In every direction, on encountering an intersection with an obstruction, simple reflected and transmitted ray directions were calculated, based on Fresnel's equations, with an effective relative permittivity of $ \varepsilon_r = 5.0 $ \cite{Seidel94a}. New source rays at each boundary were then recursively traced in the reflection and transmission directions to the next encountered obstruction on the propagating ray path. Path loss was calculated based on the CI LOS path loss model \cite{Sun16c}, with a TR separation distance equal to the total propagated ray length. Reflection and transmission losses at each obstruction were computed based on Fresnel's equations \cite{Seidel94a}. Rays were considered to contribute to the received power if they fell within the detection sphere of the RX \cite{Seidel94a}.

\begin{figure}[]
	\centering
	\includegraphics[width=0.5\textwidth]{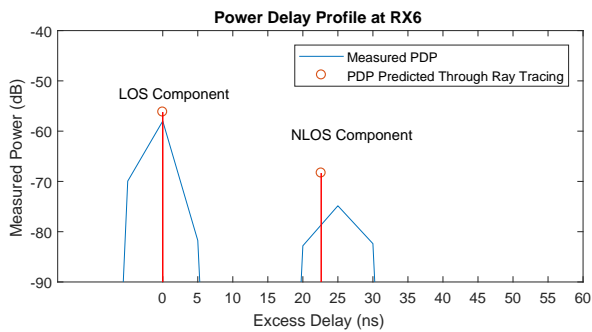}
	\caption{ Comparision of the measured PDP and the PDP predicted by ray tracing at RX 6.}
	\vspace{-15pt}
	\label{fig:PDP}
	\end{figure}
The 2-D ray tracer developed here matches the real measurements in \cite{Deng15a} to reasonable accuracy. Consider RX6, receiving signal from TX 1, as shown in Fig. \ref{fig:Indoor_locations}. Based on indoor measurements at 28 GHz \cite{Deng15a}, RX6 has a LOS peak power of -58 dBm, with a multipath component having an excess delay of 25 ns, carrying -64.8 dB power. As seen in Fig. \ref{fig:PDP}, the ray tracer predicts the LOS component to have a peak power of -56.2 dB, with the multipath component having an excess delay of 23 ns, and a power of -68.3 dB. This performance was typical for the ray tracer. The ray tracer will be used and improved upon to investigate positioning methods.

\section{Conclusion}\label{section9}
Positioning is a well-researched field with several popular positioning techniques. However, little is known of how well these positioning techniques can work in real-world mmWave systems. This paper provided an investigation of three well-known location methods - AoA-based positioning, positioning using a combination of measured path loss and AoA, and RSSI rank-based positioning.  RSSI rank vector-based positioning provided a rough estimate of the user's position. With indoor measured data at 28 GHz (and 73 GHz), and using $15^\circ$ (and $30^\circ$) HPBW rotatable antennas at the RX, we showed that using a system with a bandwidth of 800 MHz allows for good position locationing accuracy when combining AoA and path loss. The accuracy ranged from 16 cm to 3.25 m, with a mean positioning error of 1.86 m when one outlier was removed. Localizing the user using a combination of AoA and path loss worked better than localization based solely on AoA. Machine learning or data fusion, which will be implemented in further work, may provide better results. Owing to the narrow beamwidth of mmWave rays, the energy in the elevation and azimuth angle of incoming signals can be accurately estimated using ray tracing. Hence, by extending the ray tracer to 3-D and by accurately characterizing the reflection and transmission losses of obstructions, more accurate prediction of received signal characteristics and position location approaches will be possible in future works.

\section{Acknowledgments}
This material is based upon work supported by the NYU WIRELESS Industrial Affiliates Program and National Science Foundation (NSF) Grants: 1702967 and 1731290.
\bibliographystyle{IEEEtran}
\bibliography{Ojas_bib_v1}

\end{document}